\documentclass[aps,prb,reprint,superscriptaddress]{revtex4-2}
\usepackage{siunitx}
\usepackage{graphicx}			% Bilder

\usepackage{color}

\begin{document}
\newcommand{\avec}{\textbf{\textit{a}}}
\newcommand{\bvec}{\textbf{\textit{b}}}
\newcommand{\cvec}{\textbf{\textit{c}}}
\newcommand{\Hvec}{\textbf{\textit{H}}}
\newcommand{\CsFeBr}{Cs$_3$Fe$_2$Br$_9$}
\newcommand{\CsFeCl}{Cs$_3$Fe$_2$Cl$_9$}
\newcommand{\CsCrCl}{Cs$_3$Cr$_2$Cl$_9$}
\newcommand{\tfrac}[2]{{\footnotesize \frac{#1}{#2}}} % if not using amsmath
\newcommand{\boldsymbol}[1]{\textbf{\textit{#1}}} % if not using amsmath

%\title{Magnetic ordering and strong anisotropy in triangular magnet Cs$_3$Fe$_2$Br$_9$}
\title{Multiple field-induced phases in the frustrated triangular magnet Cs$_3$Fe$_2$Br$_9$} %with strong anisotropy}

\author{D.~Br\"uning}
\affiliation{$I\hspace{-.1em}I$. Physikalisches Institut, Universit\"at zu K\"oln, Z\"ulpicher Stra\ss e 77, D-50937 K\"oln, Germany}
\author{T.~Fr\"ohlich}
\affiliation{$I\hspace{-.1em}I$. Physikalisches Institut, Universit\"at zu K\"oln, Z\"ulpicher Stra\ss e 77, D-50937 K\"oln, Germany}
\author{D. Gorkov}
\affiliation{$I\hspace{-.1em}I$. Physikalisches Institut, Universit\"at zu K\"oln, Z\"ulpicher Stra\ss e 77, D-50937 K\"oln, Germany}
\affiliation{Heinz Maier-Leibnitz Zentrum (MLZ), Technische Universit\"at M\"unchen, Lichtenbergstr. 1, 85748 Garching, Germany}
\author{I.~C\'isa\v{r}ov\'a}
\affiliation{Department of Inorganic Chemistry, Charles University in Prague, Hlavova 2030/8, 128 43 Prague 2, Czech Republic}
\author{Y.~Skourski}
\affiliation{Hochfeld-Magnetlabor Dresden (HLD-EMFL), Helmholtz-Zentrum Dresden-Rossendorf, 01314 Dresden, Germany}
\author{L.~Rossi}
\affiliation{High Field Magnet Laboratory (HFML-EMFL) and Institute for Molecules and Materials, Radboud University, 6525 ED Nijmegen, The Netherlands}
\author{B.~Bryant}
\affiliation{High Field Magnet Laboratory (HFML-EMFL) and Institute for Molecules and Materials, Radboud University, 6525 ED Nijmegen, The Netherlands}
\author{S.~Wiedmann}
\affiliation{High Field Magnet Laboratory (HFML-EMFL) and Institute for Molecules and Materials, Radboud University, 6525 ED Nijmegen, The Netherlands}
\author{M. Meven}
\affiliation{RWTH Aachen University, Institut f\"ur Kristallographie, 52056 Aachen, Germany}
\affiliation{J\"ulich Centre for Neutron Science JCNS at Heinz Maier-Leibnitz Zentrum (MLZ)}
\author{A.~Ushakov}
\affiliation{M.N. Mikheev Institute of Metal Physics, Ural Branch, Russian Academy of Sciences, 620137 Ekaterinburg, Russia}
\author{S.V.~Streltsov}
\affiliation{M.N. Mikheev Institute of Metal Physics, Ural Branch, Russian Academy of Sciences, 620137 Ekaterinburg, Russia}
\affiliation{Ural Federal University, 620002 Ekaterinburg, Russia}
\author{D.~Khomskii}
\affiliation{$I\hspace{-.1em}I$. Physikalisches Institut, Universit\"at zu K\"oln, Z\"ulpicher Stra\ss e 77, D-50937 K\"oln, Germany}
\author{P. Becker}
\affiliation{Abteilung  Kristallographie,  Institut  f\"ur  Geologie  und  Mineralogie, Universit\"at zu K\"oln, Z\"ulpicher Straße  49b,  50674  K\"oln,  Germany}
\author{L. Bohat\'{y}}
\affiliation{Abteilung  Kristallographie,  Institut  f\"ur  Geologie  und  Mineralogie, Universit\"at zu K\"oln, Z\"ulpicher Straße  49b,  50674  K\"oln,  Germany}
\author{M.~Braden}
\affiliation{$I\hspace{-.1em}I$. Physikalisches Institut, Universit\"at zu K\"oln, Z\"ulpicher Stra\ss e 77, D-50937 K\"oln, Germany}
\author{T.~Lorenz}
\email{tl@ph2.uni-koeln.de}
\affiliation{$I\hspace{-.1em}I$. Physikalisches Institut, Universit\"at zu K\"oln, Z\"ulpicher Stra\ss e 77, D-50937 K\"oln, Germany}

\date{\today}

\begin{abstract}
The recently discovered material Cs$_3$Fe$_2$Br$_9$ contains Fe$_2$Br$_9$ bi-octahedra forming triangular layers with hexagonal stacking along the $c$ axis.
In contrast to isostructural Cr-based compounds, the zero-field ground state is not a nonmagnetic $S=0$ singlet-dimer state.
Instead, the Fe$_2$Br$_9$ bi-octahedra host semiclassical $S=5/2$ Fe$^{3+}$ spins with a pronounced easy-axis anisotropy along $c$ and interestingly, the intra-dimer spins are ordered ferromagnetically.
The high degree of magnetic frustration due to (various) competing intra- and inter-dimer couplings leads to a surprisingly rich magnetic phase diagram. Already the zero-field ground state is reached via an intermediate phase, and the high-field magnetization and thermal expansion data for $H\parallel c$ identify ten different ordered phases.
Among them are phases with constant magnetization of 1/3, respectively 1/2 of the saturation value, and several transitions are strongly hysteretic with pronounced length changes reflecting  strong magnetoelastic coupling.
\end{abstract}

% insert suggested PACS numbers in braces on next line
% insert suggested keywords - APS authors don't need to do this
%\keywords{}

\maketitle

\section{Introduction}

Magnetic triangular lattices show a large variety of interesting physics and have been intensively studied  \cite{Collins1997,Kawamura2001,Moessner2001,Starykh2015,Balents2010}.
For example, the search for spin-liquid candidates caused intense studies of hexagonal and triangular magnetic materials like Cs$_2$Cu$X_4$ with $X=$~Cl,~Br, Na$_2$IrO$_3$, $\alpha$-Li$_2$IrO$_3$ and Ba$_3$TiIr$_2$O$_9$ \cite{Chaloupka2010, Choi2012, Winter2017, Dey2012, Sakamoto2006}.
In contrast to the theoretical concept of a spin liquid, all these materials show magnetic order at low temperatures.
While in the 5d materials spin-orbit coupling plays an important role, the exchange interactions are typically dominating in 3d transition-metal compounds. In Cs$_3$Cr$_2X_9$ with $X=$~Cl,~Br, which contain a hexagonal arrangement of face-sharing Cr$_2X_9$ bi-octahedra, the magnetism is dominated by a strong antiferromagnetic intra-dimer coupling. This yields a singlet ground state and a field-induced magnetic ordering which can be described as a Bose-Einstein condensation (BEC) of magnons \cite{Ziman2005,Zapf2014}.
Two examples of triangular magnets with rich phase diagrams are the $S=1/2$ Heisenberg system Cs$_2$CuBr$_4$ that shows nine field-induced phase transitions and a multitude of fractional magnetization plateaus \cite{Fortune2009}, and the semi-classical $S=5/2$ material RbFe(MoO$_4$)$_2$ with five ordered phases \cite{Smirnov2007}. In these triangular lattices, magnetic moments lie within the plane and the interesting phase diagrams occur for in-plane applied fields. Another example is CuFeO$_2$, where the Fe$^{3+}$ moments order perpendicular to the triangular planes~\cite{Mitsuda1991,Inosov2019}.

Recently, the new material \CsFeBr\ was discovered that is isostructural to Cs$_3$Cr$_2$X$_9$ and crystallizes in the hexagonal space group $P$6$_3$/$mmc$ with $a=\SI{7.5427(8)}{\angstrom}$ and $c=\SI{18.5849(13)}{\angstrom}$.\cite{Wei2018} The structure consists of face-sharing octahedra forming Fe$_2$Br$_9$ bi-octahedra in triangular layers.
The shortest Fe-Fe distance amounts to \SI{3.585(3)}{\angstrom} within the bi-octahedra. The in-plane Fe-Fe$_{\text{p}}$ distance within the triangular layers is about twice as large, \SI{7.179(2)}{\angstrom}, and the interlayer Fe-Fe$_{\text{c}}$ distance of \SI{7.543(1)}{\angstrom} is slightly larger. Based on a study of powder samples, a band gap of \SI{1.65}{\electronvolt} and antiferromagnetic order at $T_\text{N}=\SI{13}{\kelvin}$ were reported \cite{Wei2018}. Here, we present a detailed study of the low-temperature ordered phases of \CsFeBr\ single crystals up to the saturation magnetization $M_\text{S}$ that is reached at 43\,T (52\,T) for a field parallel (perpendicular) to the \cvec\ axis.

\begin{figure*}
	\includegraphics[width=17cm]{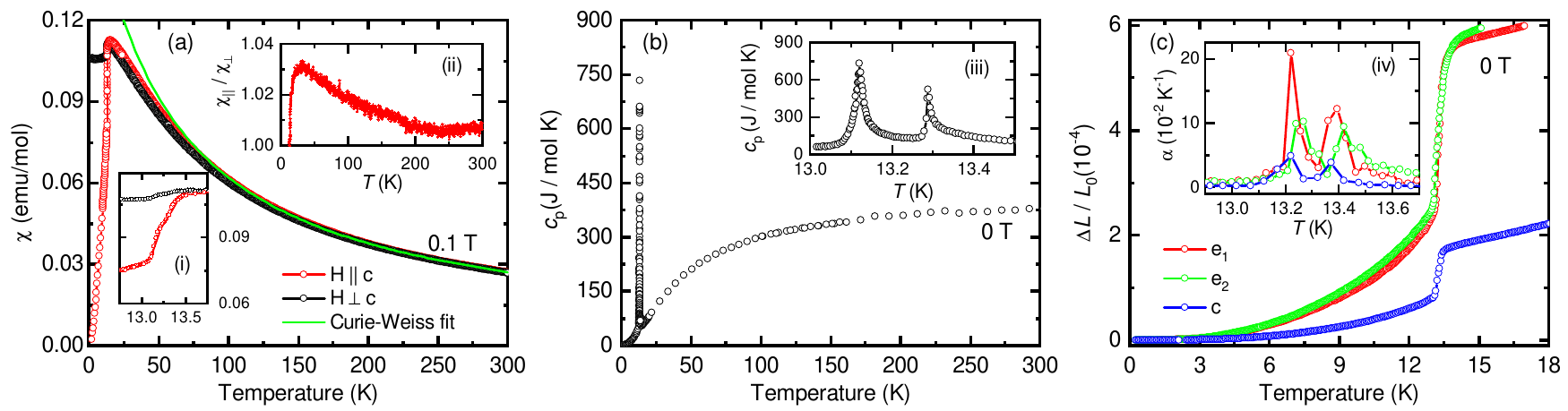}
	\caption{(a) Temperature dependence of the magnetic susceptibility $\chi_i$ for \Hvec$\parallel$\cvec\ (red) and \Hvec$\perp$\cvec\ (black). The green line stems from a Curie-Weiss analysis of $\chi_\parallel$ for $T > \SI{100}{\kelvin}$. t (i) An expanded view around the two-step ordering transition, and (ii) $\chi_\parallel/\chi_\perp$ revealing an essentially isotropic susceptibility in the high-temperature paramagnetic phase. (b) Zero-field specific heat with a huge anomaly around 13.2\,K arising from the magnetic order, which actually evolves via a two-step transition with $T_{\text N1}= \SI{13.29}{\kelvin}$ and $T_{\text N2}= \SI{13.12}{\kelvin}$ as shown in (iii). (c) Thermal expansion in \SI{0}{\tesla} measured along the hexagonal $c$ axis and along two perpendicular directions $e_1$ and $e_2$ within the $ab$ plane. (iii) The corresponding uniaxial thermal expansion coefficients $\alpha= 1/L_0\cdot\partial \Delta L_i / \partial T$ again signal the two-step transition.}
	\label{chi_cp_dL}
\end{figure*}

\section{Experimental}

Single-crystals were grown from aqueous solutions of CsCl and FeBr$_3$ in a molar ratio of $\sim$2:1 and a surplus of HBr at room temperature during a period of six months. Further studies revealed that \CsFeBr\ can be grown from solutions of CsBr and FeBr$_3$ in the range of 3:2 and 4:1 between room temperature and \SI{50}{\degreeCelsius}.

Using commercial setups (Quantum Design PPMS and MPMS), specific heat and magnetic susceptibility were measured between \SIrange{2}{300}{\kelvin}, and the low-temperature magnetization was measured up to 14\,T. High-field magnetization data were obtained using pick-up coils in pulsed magnetic fields up to \SI{56}{\tesla} at the high-field center HLD, Dresden Rossendorf.
Thermal expansion and magnetostriction $\Delta L(T,H)/L_0$ were measured in a home-built capacitance dilatometer down to \SI{0.26}{\kelvin} in magnetic fields up to \SI{17}{\tesla}~\cite{Lorenz2007,Bohaty2013}. The field was applied parallel to the crystal direction whose length change $\Delta L_i$ was measured. $L_0$ denotes the corresponding overall length of the sample, and the uniaxial thermal expansion coefficient $\alpha = 1/L_0\cdot\partial\Delta L_i/\partial T$ was obtained numerically. Additonal high-field expansion data up to \SI{37}{\tesla} were taken at HFML Nijmegen using commercial dilatometers \cite{Kuchler2012,Kuchler2017a}.

The crystal structure of \CsFeBr\ was investigated via an APEX (Bruker) four-circle single crystal X-ray diffractometer at $ \SI{150}{\kelvin}$.~\cite{CSD}
The low-temperature crystal and magnetic structure was studied on the single-crystal neutron diffractometer HEiDi \cite{Meven2015} and on the KOMPASS instrument (both at FRM-II, Munich).
On HEiDi, a crystal of $ \SI{48.680(14)}{\milli\gram} $ was mounted in a way that the $ (0\,1\,1) $ direction was oriented along the $ \varphi $ axis of the four circle diffractometer and data were collected with wavelengths $\lambda$ of $ \SI{1.171}{\angstrom} $ and $ \SI{0.795}{\angstrom} $. On KOMPASS, the measurements were performed in the (100)/(010) and (100)/(001) scattering planes. A polarized beam was obtained through
serial polarizing V-shaped multichannel cavities and a highly oriented pyrolytic graphite (HOPG(002)) monochromator, $\lambda=\SI{4}{\angstrom}$.
An additional V-shaped multichannel cavity was used to analyze the polarization of the scattered beam in the experiments with the second scattering plane yielding a flipping ratio of 11. Higher order contamination was suppressed with a velocity selector.

\section{Results and discussion}

Figure \ref{chi_cp_dL}(a) shows the temperature dependent magnetic susceptibility for \Hvec\ $\parallel$ \cvec\ and  \Hvec\ $\perp$ \cvec. In the high-temperature regime, $\chi_i$ is isotropic and well described by a Curie-Weiss law.
Fixing the spin of the Fe$^{3+}$ ions to $S=5/2$ the
Curie-Weiss analysis of $\chi_{\parallel c}$ yields a Curie temperature $\theta=-\SI{56}{\kelvin}$ and a reasonable $g$-factor of $2.09$.\cite{chi_PM} Below \SI{13}{\kelvin}, $\chi_{\parallel c}$ drops to zero while $\chi_{\perp c}$ remains almost constant suggesting nearly collinear antiferromagnetic order with moments pointing essentially along \cvec. The magnetic system is weakly frustrated as is indicated by the frustration factor $|\theta|/T_\text{N}\approx4.4$.
Figure \ref{chi_cp_dL}(b) shows the specific heat, which reveals a huge anomaly around \SI{13.2}{\kelvin}. As shown in the inset (iii) and discussed in Appendix~\ref{app:HC}, we resolve two distinct transitions at $T_\text{N1}=$ \SI{13.29}{\kelvin} and $T_\text{N2}=$ \SI{13.12}{\kelvin}. The two-step transition is also visible in the expanded view (i) of the susceptibility.

Figure~\ref{chi_cp_dL}(c) shows thermal expansion data measured along the hexagonal \cvec\ axis and along the two in-plane directions $e_{1}$ and $e_{2}$, which are parallel and perpendicular to the hexagonal in-plane axes \avec, respectively.
At the transition each data set shows a step-like relative length change with $\Delta L_c/L_0 \simeq 10^{-4}$ (blue) and $\Delta L_{e_i}/L_{0} \simeq 3 \cdot 10^{-4}$ for both, $e_1\|\avec$ (red) and $e_2\perp\avec$ (green). The enlarged view on the thermal expansion coefficients $\alpha$ in the inset (iv) reveals that each of them also shows two distinct peaks \footnote{The thermal-expansion anomalies are shifted by $\simeq 100\,$mK compared to the $c_p$ anomalies, which exceeds the typical experimental uncertainties arising from the usage of different samples in different experimental setups. This difference partly arises from the fact that the $c_p$ anomalies stem from the long-pulse relaxation curves, see Appendix~\ref{app:HC}, whereas $\alpha$ was measured upon heating. In addition, a weak upwards shift of the transition temperatures can arise due to the uniaxial pressure which is applied when the sample is clamped into the capacitance dilatometer.}. As will be discussed below, the magnetic transition breaks the 3-fold in-plane symmetry and allows for three twin domains of orthorhombic symmetry. Thus, a finite magnetoelastic coupling should induce different thermal expansion anomalies of $e_1$ and $e_2$ in a single-domain sample.  However, in the capacitance dilatometer the crystal is fixed by CuBe springs, which apply a weak uniaxial pressure along the measured $\Delta L_i$ and may cause a partial or full detwinning at the symmetry-breaking phase transition~\cite{Niesen2013, Niesen2014a}. In this case, the $\Delta L_i/L_0$ measured along $e_1$ and $e_2$
result from different orientations of twin domains, as is discussed in the Appendix~\ref{app:twins}.

\begin{figure}
	\includegraphics[width=8.1cm]{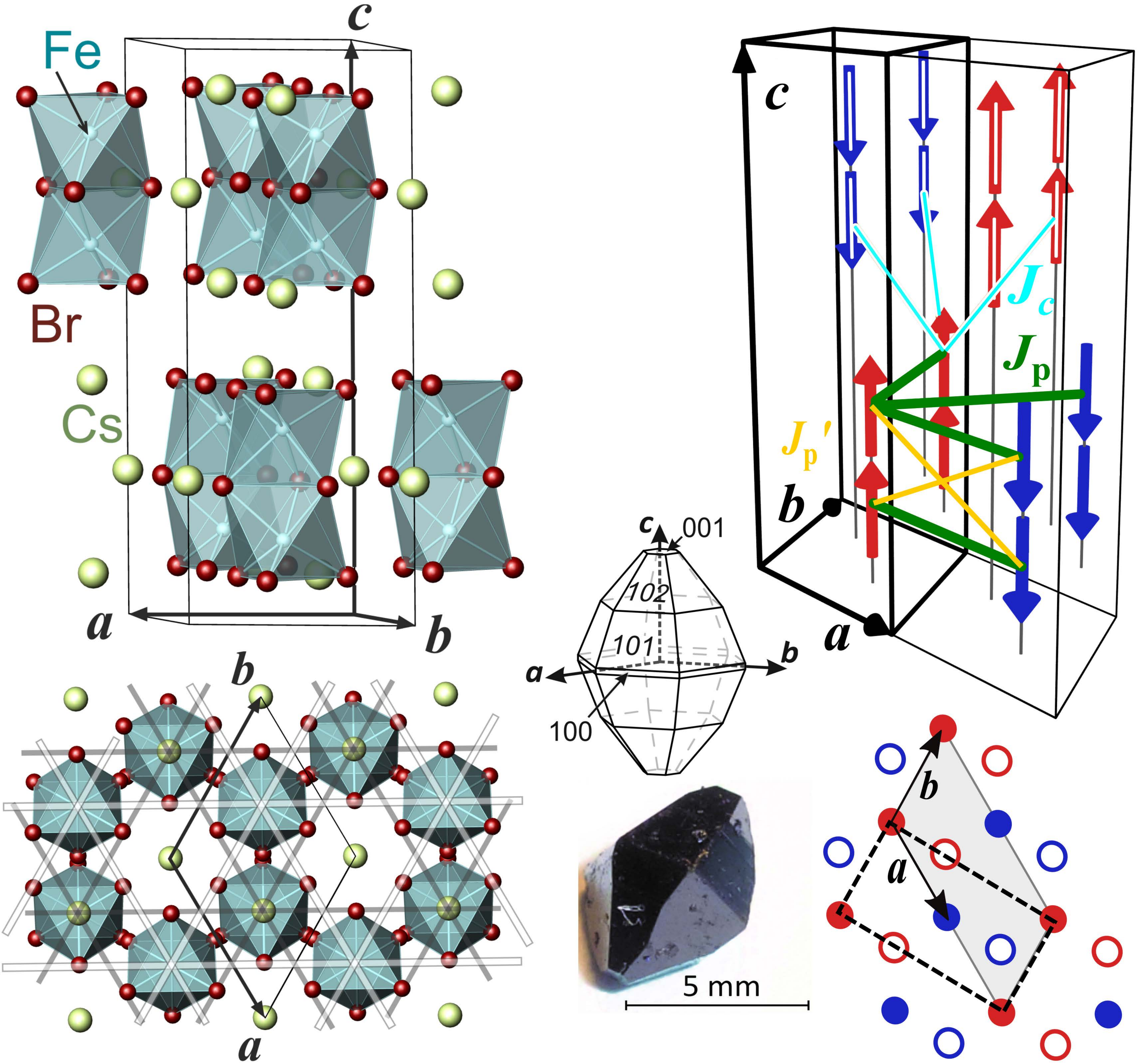}
	\caption{\label{magnetic_structure} Crystal (left) and magnetic (right) structure of \CsFeBr . The top panels are 3D versions and the bottom panels show the respective projections to the $ab$ plane. In the lower left panel, the solid and open lines connect the Fe$_2$Br$_9$ bi-octahedra of the lower- and the upper triangular planes, respectively. The magnetic structure was determined for the zero-field phase P1 at $ \SI{2.5}{\kelvin}$, which shows essentially parallel Fe spins within the bi-octahedra. The magnetic unit cell leaves \cvec\ unchanged, while the hexagonal in-plane axis $\avec$ is doubled, which results in an orthorhombic unit cell. This is illustrated by the dashed rectangle in the projection to the $ab$ plane, where the open (filled) circles mark Fe ions from the upper (lower) Fe$_2$Br$_9$ layer, with their magnetic moments pointing either along +\cvec\ (red) or -\cvec\ (blue). Besides the intradimer interaction $J$ (not shown) we discuss the interdimer interactions $J_p$ (green) and $J_p'$ (yellow) acting on nearest-neighbor dimers in the same layer, and the interlayer interaction $J_c$ (blue) coupling nearest-neighbor spins in adjacent dimer layers. The photo shows a single crystal with pronounced facets reflecting the hexagonal structure.}
\end{figure}

The crystal structure of \CsFeBr\ is displayed in Fig.~\ref{magnetic_structure}. The fundamental building blocks are Fe$_2$Br$_9$ bi-octahedra forming triangular planes, which are arranged in the usual $ABAB$ stacking of the hexagonal crystal structure. In the $ab$ projection of the crystal structure (Fig. 2 lower left panel), the bi-octahedra of the different triangular planes are connected by open and solid gray lines. The right panel of Fig.~\ref{magnetic_structure} shows the Fe$^{3+}$ magnetic moments, which are aligned parallel to each other within the Fe$_2$Br$_9$ dimers for the zero-field ordered phase (see below), and the most important interdimer  magnetic exchange interactions are also indicated. Neglecting the interlayer coupling $J_c$, the bi-octahedron dimers form frustrated triangular planes which have been long studied~\cite{Wannier1950}. The triangular arrangement is depicted by using open (closed) symbols for the Fe spins of the upper (lower) planes of the 3D unit cell in Fig.~\ref{magnetic_structure}. When adding the interlayer coupling, $J_c$, on an equal footing to $J_p$, and when ignoring the intradimer coupling $J$ as well as $J_p'$, one can also argue about a realization of staggered honeycomb magnetic planes. Note that for antiferromagnetic $J_p$ and $J_c$ both are frustrated, but $J_p$ couples to six neighboring spins of the same triangular plane, while $J_c$ couples to three neighbors of the next triangular plane.

%% In this figure I wpoold  make more clear that we actually have triangular layers, not honeycomb. One can for example make bioctahedra in different layers in (a) above and especially below by different colors: e.g. lower layer,  A in the stacking, make blue as now, and the upper layer, B in stacking - say by pink. Then especially in the  lower figure (a) it would be clear that it is not honeycomb. Another option is to add figure of the type I send you, on my slide 1, showing lower triangilar layer by blue lines/bonds, the upper one - by red ones, and the "vertical" bonds between lower and upper bioctahedra - by weaker, say dashed lines (not solid green lines as in my figure - in my figure these weak interlayer bonds look too prominent; I would make them weaker.

\begin{figure}[t]
	\includegraphics[width=0.7\columnwidth]{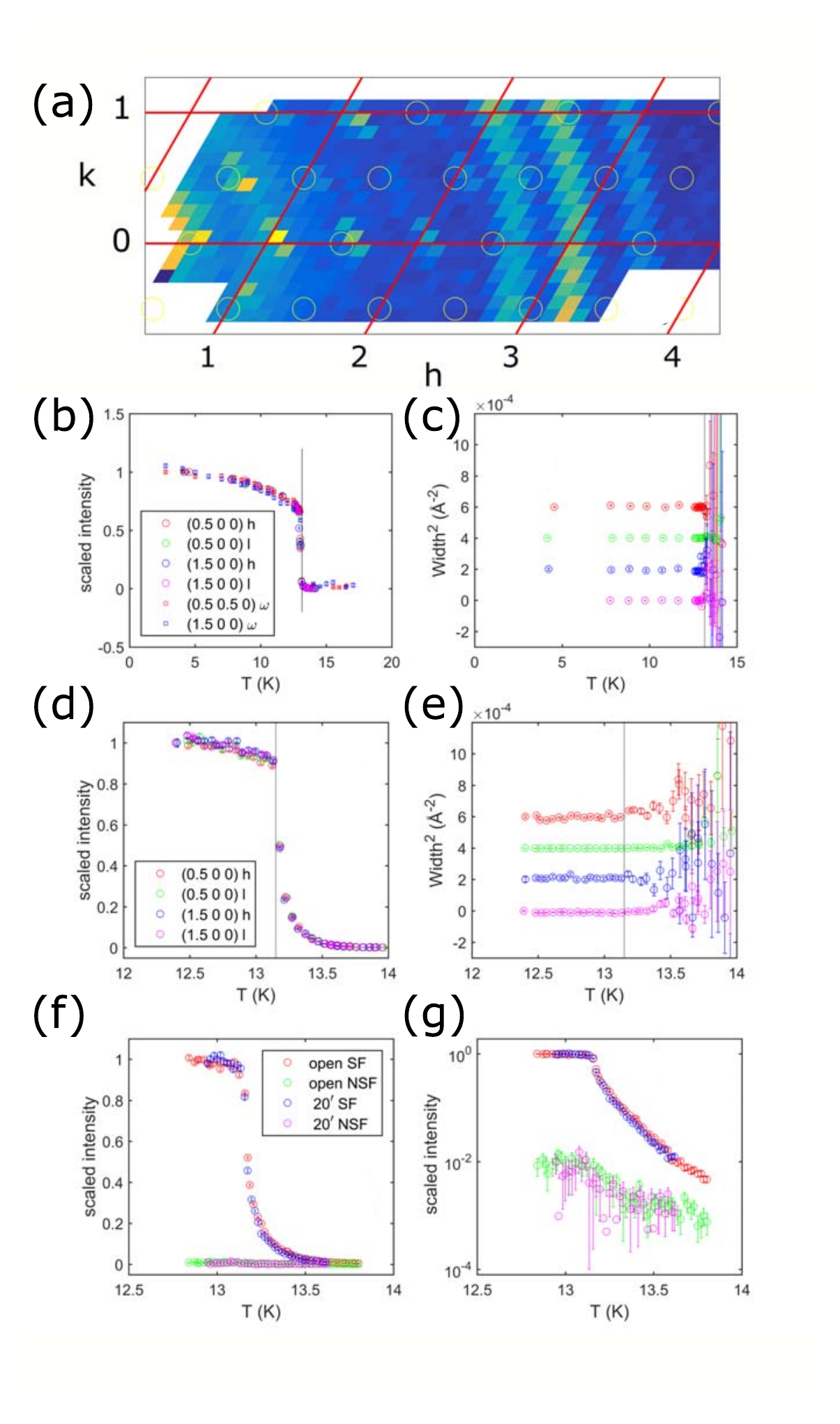}
	%\vskip-1cm
	\caption{(a) Neutron diffraction intensity in the reciprocal $(h\,k\,l\,$=$\,0)$ plane of \CsFeBr\
		observed on HEiDi at 2.5\,K in zero field. Magnetic superstructure reflections ($\frac{3}{2}$,0,0), (1,$\frac{1}{2}$,0), and ($\frac{1}{2}$,$\frac{1}{2}$,0) as well as others are clearly visible. Nuclear Bragg reflections exist at the crossings of the red lines, and magnetic peaks are indicated by yellow circles.
		(b) Peak amplitudes and (c) quadratic widths obtained by fitting Gaussian profiles to the $h$, $k$ (KOMPASS) or rocking scans (HEiDi) across magnetic peaks. Only for the KOMPASS data quadratic widths are given after subtraction of the low-temperature values (data are vertically shifted for clarity). (d,e) The same analysis for data obtained with polarization analysis (SF with neutron polarization axis along the vertical direction) that focus
on the temperature range close to the transition.  (f,g) Intensities at the magnetic Bragg positions as a function of temperature for SF and nSF channels (measurement with open and 20$^\prime$ collimation before the sample). All data were taken upon heating.}
	\label{neutron}
\end{figure}

The crystal structure was refined in space group $ P6_3/mmc $ at 150, 15, and 2.5\,K yielding no significant differences between the two lowest temperatures and only slight changes in the Br positions when comparing 150\,K and low-temperature structures, see Appendix~\ref{app:struc}. At 2.5\,K, {\it i.e.} well in the magnetically ordered phase P1, we searched for magnetic Bragg peaks. As is shown in Fig.~\ref{neutron}(a), a mapping of the $(hk0)$ plane in reciprocal space yields additional magnetic Bragg peaks at various half-indexed scattering vectors of type ($\frac{n+1}{2}$,$k$,0) determining the magnetic propagation vector to ($\frac{1}{2}$,0,0). The star of this propagation vector in the hexagonal structure contains ($\frac{1}{2}$,0,0), (0,$\frac{1}{2}$,0) and (-$\frac{1}{2}$,$\frac{1}{2}$,0) as can be seen in Fig.~\ref{neutron}(a).
All observed magnetic Bragg peaks can be indexed with one of these three propagation vectors. We also looked for magnetic Bragg peaks appearing at a half-integer $l$ component but did not observe such intensities. The antiferromagnetic order with
these propagation vectors corresponds to the stripe order in a single triangular layer that is illustrated in the lower right part of Fig.~\ref{magnetic_structure}. The different domains correspond to the propagation vectors and to the stripes rotated by 0, 60 and 120 degrees.

Temperature dependent neutron diffraction experiments were performed on HEiDi and on KOMPASS. Fig.~\ref{neutron}(b) and (c) show the
amplitudes and the squares of the widths obtained by fitting Gaussian profiles to the scans.
The half-indexed magnetic Bragg intensities show only moderate temperature dependence in the ordered phase and abruptly disappear at the lower transition temperature $T_{N2}$ in accordance with the first-order character of this phase transition. The scan width only increases above $T_{N2}$.
In order to more deeply study the phase transition additional experiments using polarization analysis were performed on KOMPASS, Fig.~\ref{neutron}(d) and (e).
These experiments reveal sizeable diffuse scattering above $T_{N2}$ that persists also well above $T_{N1}$. At the latter, higher transition
temperature we find no signature in the temperature dependence of the diffuse scattering at the half-indexed scattering vector positions.
For the two reflections studied, ($\frac{1}{2}$,0,0) and ($\frac{3}{2}$,0,0), scans were performed in $h$ and in $l$ directions so that the correlations parallel and perpendicular
to the planes can be compared. First, the lower transition is not associated with an abrupt increase in the widths, which points to the first-order character
and partial coexistence with another ordering scheme in phase P2. Second, there is no indication
for a qualitatively distinct behavior when scanning parallel and perpendicular to the planes.
In contrast, for a quasi-two dimensional system one would expect two-dimensional correlations to survive
above the N\'eel temperature while the correlation perpendicular to the planes will broaden more rapidly.
Therefore, Cs$_3$Fe$_2$Br$_9$ is essentially a three-dimensional magnetic system.
With the polarization analysis, one may directly determine the orientation of the magnetic moments.
We set the neutron polarization axis perpendicular to the scattering plane defined by (1,0,0) and (0,0,1), therefore the spin-flip (SF) channel at ($\frac{1}{2}$,0,0) and ($\frac{3}{2}$,0,0)
records magnetic contributions parallel to $c$ while the non-spin-flip (nSF) channel detects contributions parallel to $b$.
At the half-indexed position there are no nuclear contributions.
The data in Fig.~\ref{neutron}(f) and (g) are corrected for the finite flipping ratio and unambiguously reveal that the magnetic intensities
at the half-indexed reflections completely arise from moments pointing along $c$, which agrees with the vanishing magnetic susceptibility
$\chi_{\perp c}(T\rightarrow 0)$, Fig~\ref{chi_cp_dL}(a).
The diffuse scattering visible between $T_{\text{N2}}$ and $T_{\text{N1}}$ and above $T_{\text{N1}}$ exhibits the same magnetic anisotropy so that also
the short-range correlations are associated with moments along $c$.

The symmetry analysis for the ($\frac{1}{2}$,0,0) propagation vector in space group $P6_3/mmc$ was performed with the FullProf program package~\cite{Rodriguez-Carvajal1993} and is discussed in the Appendix~\ref{app:mag}. The four Fe ions in the primitive cell are all equivalent in the magnetic phase. Furthermore, susceptibility and neutron polarization
analysis reveal that the essential part of the magnetic moment aligns along the $c$ direction. Only $\Gamma_2$, $\Gamma_3$, $\Gamma_6$, and $\Gamma_7$ possess
a finite $c$ component, so that the other magnetic models can be excluded.
The four irreducible representations correspond to antiferromagnetic or ferromagnetic dimers (two spins
in the bi-octahedron) combined with a ferro- or antiferromagnetic stacking within the unit cell. Refinements were performed with these four magnetic structures
taking the three domain orientations into account.
The data is only compatible with $\Gamma_3$ yielding a weighted $R$ value of 7.7\%, while 77, 73 and 80\% are obtained for $\Gamma_2$, $\Gamma_6$, and $\Gamma_7$, respectively. The parallel alignment of the moments within a dimer can already be deduced from the fact that the strongest magnetic peaks are found in the $(hk0)$ plane, while for an antiferromagnetic alignment these intensities exactly cancel (because the two Fe ions exhibit the same $x$ and $y$ coordinates).
Also the antiferromagnetic stacking of the spins within the cell that arises through $J_c$ is unambiguous.

The symmetry analysis indicates that the $c$ moments in $\Gamma_3$ can be accompanied by an in-plane moment, {\it i.e.} a weak canting. Within a dimer the ferromagnetic
$c$ moments are coupled with antiferromagnetic in-plane moments arising from Dzyaloshinski-Moriya interaction. The magnetic refinement with the Fullprof package
only slightly improve with the in-plane moment yielding a total moment of 3.954(5)\,$\mu_\text{B}$, a moment along $c$ of  3.925(5)\,$\mu_\text{B}$
and an in-plane component of 0.47(14)\,$\mu_B$. The three domains occupy similar volume fractions of 39, 30 and 31\%.
The size of the in-plane moment is consistent with the small reduction of the in-plane susceptibility
in the ordered phase. The magnetic structure is illustrated in Fig.~\ref{magnetic_structure} neglecting the in-plane component.
Here, the spin directions up and down are marked in red and blue, respectively, while spins from the upper (lower) Fe$^{3+}$ double-layer are depicted by open (closed) symbols. Only two thirds of the in-plane nearest-neighbor (NN) spins show antiparallel orientations and, analogously, only two thirds of the NN spins across the neighboring planes of the $AB$-stacked bi-octahedra are antiparallel to each other. Thus, all the NN inter-dimer couplings $J_p$, $J_p'$, and $ J_c$ are geometrically frustrated.

The stripe order described by the ($\frac{1}{2}$,0,0) propagation vector is one possible lowest-energy arrangement of the triangular frustrated magnetic
lattice \cite{Wannier1950}. It breaks rotational symmetry and the degeneracy can be lifted by magnetoelastic coupling, as it is discussed in Appendix~\ref{app:twins}.
The magnetic structure and the magnetoelastically distorted structure can be described in the orthorhombic space group $ Cmcm $ (No. 63) which results from $P6_3/mmc$ by breaking the three-fold axis. A refinement of the additional structural parameters in the lower space group improves the R values \cite{Karplus2012} only slightly
\footnote{Improvement by lower space group: From $ R(\mathrm{obs}) = 3.35\,\%, \mathrm{w}R(\mathrm{obs}) = 3.53\,\%, R(\mathrm{all}) = 6.15\,\%, \mathrm{w}R(\mathrm{all}) = 3.81\,\% $ to $ R(\mathrm{obs}) = 3.26\,\%, \mathrm{w}R(\mathrm{obs}) = 3.43\,\%, R(\mathrm{all}) = 5.97\,\%, \mathrm{w}R(\mathrm{all}) = 3.63\,\% $ for the data measured at $ \SI{2.5}{\kelvin} $} and a similar improvement can be achieved for the data measured at $ \SI{15}{\kelvin} $, so the structural symmetry reduction induced by the magnetoelastic coupling cannot be resolved in the neutron diffraction study.

\begin{figure*}
	\includegraphics{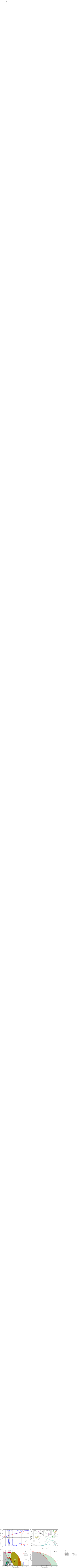}
	\caption{(a) Pulsed-field magnetization and the differential susceptibility measured at \SI{1.5}{\kelvin} up to \SI{55}{\tesla}. The data for \Hvec\ $\parallel$ \cvec\ (blue) reveal 7 phase transitions and 2 transitions for \Hvec\ $\perp$ \cvec\ (red), as marked by the vertical lines. Note that there are 2 plateau phases (P5 and P6) with constant magnetization of, respectively, 1/3 and 1/2 of the saturation $M_\text{S}$. (b) Magnetostriction curves $\Delta L_c(H)/L_0$ obtained with increasing \Hvec\ $\parallel$ \cvec\ up to \SI{15}{\tesla} for representative temperatures from 0.26 to \SI{10}{\kelvin}, (i) contains an additional curve up to \SI{37}{\tesla}. The kinks or jump-like changes in $\Delta L_c(H)$ signal, respectively, 2$^{nd}$- or 1$^{st}$-order phase transitions with 10 ordered phases (P1 to P10) and the paramagnetic (PM) state. The existence regions of the various phases are marked via the changing colors of $\Delta L_c(H,T)$. Strong hysteresis effects occur at \SI{0.26}{\kelvin} as is shown in (ii). (iii, iv) Corresponding thermal-expansion $\Delta L_c(T)$ curves in the field range from 15.5 to \SI{19.5}{\tesla}.
	(c) Phase diagram for \Hvec\ $\parallel$ \cvec\ and (d) for \Hvec\ $\perp$ \cvec ; the phases are colored as in (b) and the vertical lines mark the critical fields $H_{C1-7}$ as in (a).}
	\label{MvH_dL_Phadis}
\end{figure*}

As discussed in the Appendix~\ref{app:DFT}, we mapped the DFT+U total energies of several magnetic configurations onto the Heisenberg model written as $\sum_{i>j} J_{ij} {\bf S}_i {\bf S}_j$, and from these calculations the intra- and the interdimer couplings were determined, see Fig.~\ref{magnetic_structure}. The intradimer $J$ results from the Fe-Br-Fe exchange paths via the common Br ions of the face-sharing bi-octahedron with bond angles of \ang{83.5} and is found to be weakly ferromagnetic with $J=-1\,$K and $-1.4\,$K for $U=5.8\,$eV and $6.5\,$eV, respectively. Interestingly, this coupling is smaller (by absolute magnitude) than the interdimer couplings $J_p$ and $J_c$, which both result from two Fe-Br-Br-Fe exchange paths. Here, we obtain  values $J_p = 2.8 (3.2)\,$K and $J_c = 3.0 (2.8)\,$K for U$=5.8 (6.5)\,$eV, whereas, for both U, a significantly smaller $J_\text{p}'= 0.2\,$K is obtained for the diagonal interdimer couplings which result from single Fe-Br-Br-Fe exchange paths. These parameters are in perfect agreement with the magnetic structure determined at zero field and 2.5\,K.
Based on these couplings, one can consider \CsFeBr\ as a system of frustrated  $S=5/2$ triangular layers, which are moderately coupled along \cvec. Due to the hexagonal $AABB$ stacking of the single Fe$^{3+}$ layers, the interlayer coupling alternates between the weakly ferromagnetic $J$, for $AA$ and $BB$, and the larger, but frustrated  antiferromagnetic $J_c$ for $AB$. On a mean-field level, the Weiss temperature is given by
\begin{equation}\label{key}
\theta = -\sum_i z_i J_i S(S+1)/3 \simeq -2.92 \sum_i z_i J_i
% (S(S+1)) (J+6J_p+6J_p'+3J_c)
\end{equation}
with $S=5/2$ and the coordination numbers $z_i=1$, 6, 6, and 3 for the couplings $J$, $J_p$, $J_p'$, and $J_c$, respectively. Depending on U, the ab-initio values yield $\theta \simeq -76 (-80)\,$K. In view of the fact that mean-field theory typically overestimates $|\theta|$ and only near-neighbor couplings are considered, these values well agree to the experimentally observed $\theta \simeq -56\,$K. Due to the inherent frustration of the triangular arrangement of the Fe spins, one may expect that relatively small magnetic fields can already induce variations of the magnetic structure.

Pulsed high-field magnetization data of \CsFeBr\ at \SI{1.5}{\kelvin} are shown in Fig.~\ref{MvH_dL_Phadis}(a). For \Hvec\ $\perp$ \cvec, $M(H)$ is almost linear up to the saturation
field of \SI{52}{\tesla}, and the saturation magnetization $M_\text{S}$ of 10\,$\mu_\text{B}$/fu agrees well with two $S=5/2$ Fe$^{3+}$ ions per formula unit. The differential susceptibility $\chi=\partial M/\partial H$ reveals two peaks at \SI{39}{\tesla} and \SI{52}{\tesla} indicating phase transitions. Based on additional data, we derive the  phase diagram in Fig.~\ref{MvH_dL_Phadis}(d) containing two ordered phases P1 and P2, which continuously evolve from the two zero-field transitions.

In contrast, multiple
transitions occur for \Hvec\ $\parallel$ \cvec . Both,  $\chi_\parallel$ and $M(H)$ remain almost zero up to $H_{C1}=\SI{5.6}{\tesla}$, where $M(H)$ starts to increase linearly up to $H_{C2}=\SI{12}{\tesla}$. At $H_{C2}$, $\chi_\parallel$ shows a peak corresponding to a step-like increase of $M(H_{C2})$ followed by another region of constant $\chi_\parallel$. At $H_{C3}=\SI{14}{\tesla}$, $M$ shows another step to an approximately constant $M(H>H_{C3}) \simeq 3.33\;\mu_\text{B}$/fu which equals 1/3 of $M_\text{S}$.
Another step-like increase occurs at $H_{C4}=\SI{20}{\tesla}$ followed by a wide plateau with $M\simeq 1/2 M_\text{S}$ up to $H_{C5}\approx 32\,$T.
Above this field, $\chi_\parallel$ becomes significantly enhanced again up to $H_{C7}\approx 41.5\,$T with an intermediate peak at $H_{C6}= 38\,$T and $M_\text{S}$ is finally reached above about 43\,T. This yields an easy-axis anisotropy $\delta=H_\text{an}/H_\text{ex}=2(H_\text{S}^\perp-H_\text{S}^\parallel)/(H_\text{S}^\perp+H_\text{S}^\parallel)\approx 0.2$, which is a rather large value for a $S=5/2$ material. For comparison, $\delta\approx0.1$ is found for RbFe(MoO$_4$)$_2$~\cite{Smirnov2007}, and for CuFeO$_2$ an almost isotropic  $\delta\approx0.02$ is reported~\cite{Lummen2009b,Zuo2015}. Note that from our ab-initio values of the exchange parameters a saturation field $H_\text{S}=4((J_p+J_p')+2J_c)k_{\rm B}S/g\mu_{\rm B}\simeq 33\,$T would be expected. This value is below the experimental results for both field directions, indicating that apart from an anisotropy term additional exchange couplings between more distant spins should be taken into account.

Figure~\ref{MvH_dL_Phadis}(b) summarizes representative expansion data $\Delta L_c(T,H\parallel c)$ measured in static fields up to \SI{37}{\tesla}. At \SI{0.26}{\kelvin}, there is a kink in $\Delta L_c (H_{C1}=\SI{5.2}{\tesla})$ signaling a second-order phase transition. At $H_{C2}=\SI{11.2}{\tesla}$, the length increases discontinuously by about $1.7\cdot$10$^{-4}$ followed by another, slightly smaller discontinuity at $H_{C3}=\SI{13.2}{\tesla}$. The inset (ii) resolves the pronounced hysteresis between the field-increasing and the field-decreasing run. Because the respective critical fields are shifted by about \SI{3}{\tesla}, the hysteresis regions of the two field-induced transitions overlap and, consequently, around 10.6\,T and 0.26\,K, each of the three phases can be realized depending on the field-sweep protocol. Typical first-order solid-state transitions are quasi-discontinuous, resulting, e.g., in more or less S-shaped length changes due to a finite transition width and/or phase coexistence. In contrast, the low-temperature transitions in \CsFeBr\ are extraordinarily sharp with discontinuous relative length changes of about  $10^{-4}$, which systematically change towards continuous variations in $\Delta L_c(H)$ upon increasing temperature.
The insets (i), (iii), and (iv) display further magnetostriction $\Delta L_c(H)$ and thermal expansion $\Delta L_c(T)$ measurements which signal different sequences of field- or temperature-induced magnetoelastic transitions in different regions of the phase diagram.

Combining all anomalies of the thermal-expansion, magnetostriction and magnetization data reveals the phase diagram in Fig.~\ref{MvH_dL_Phadis}(c). In zero field, there is a two-step transition with an intermediate phase P2 between the paramagnetic phase and the ground state P1. Below \SI{5}{\kelvin}, P1 shows a second-order transition to P3 at 5 to \SI{7}{\tesla}, which is followed by a cascade of very sharp first-order transitions around 9, 13, and \SI{20}{\tesla} to the phases P4, P5, and P6, respectively. Below \SI{4}{\kelvin}, these first-order transitions become strongly hysteretic. On further increasing field, phase P7 is reached through a second-order transition at \SI{32}{\tesla} followed by another first-order transition to P8 around 38\,T, and $M_\text{S}$ is finally reached at \SI{43}{\tesla}. The phases P1, P5, and P6 are characterized by essentially constant magnetization plateau values of $M=0$, 1/3~$M_\text{S}$, and 1/2~$M_\text{S}$, respectively, while the other low-temperature phases P3, P4, P7, and P8 show more or less linear $M(H)$ behavior with similar slopes $\chi_\parallel$. Above $\sim$\SI{7}{\kelvin} and below ${\sim}$\SI{18}{\tesla}, three other phases are stabilized. The intermediate phase P2, which covers only a small temperature interval of 0.2\,K between $T_\text{N1}$ and $T_\text{N2}$ in zero field, continuously grows with increasing field until it finally dominates the intermediate field range from about \SIrange{8}{18}{\tesla} at elevated temperature. The additional phases P9 and P10 only form comparatively small pockets. Phase P9 separates P2 from the low-temperature 1/2~$M_\text{S}$ plateau phase P5, and P10 is located between P2 and the high-temperature paramagnetic phase from \SIrange{10}{19}{\tesla}.

In a first attempt, we consider \CsFeBr\ with $S=5/2$ as stacked triangular layers of classical spins, which allows us to  compare our data to numerical studies of the field-temperature phase diagram obtained via Monte-Carlo simulations~\cite{Seabra2010,Seabra2011a}. The simulations considered triangular layers with antiferromagnetic in-plane NN and NNN interactions $J_{1/2}$ for a moderate frustration ratio $J_2/J_1=0.15$ that is sufficient to suppress the so-called 120$^\circ$ zero-field groundstate of pure Heisenberg spins~\cite{Loison1994}. Along \cvec , a simple $AA$ stacking with  ferromagnetic interlayer coupling $J_{\perp}/J_1=-0.15$ was kept constant, while the single-ion anisotropy energy $-DS_z^2$ was varied from zero up to the strong Ising case $D/J_1=1.5$. Interestingly, the simulation for $D/J_1=0.5$~\cite{Seabra2010}
reproduces several basic aspects of the experimental phase diagram of \CsFeBr\ surprisingly well. The obtained zero-field groundstate corresponds to P1, and a first field-induced transition of second-order is expected at $h_{c1} \simeq h_{S}/8$, in agreement with the transition from P1 to P3 at $H_{C1}=\,$\SI{5.2}{\tesla}. Moreover, fractional magnetization-plateau states with 1/3~$M_\text{S}$ and 1/2~$M_\text{S}$ are predicted, which are entered via first-order transitions, as it is observed for the phases P5 and P6. Upon increasing field, the plateau phases are expected to alternate with intermediate states of finite $\chi_\parallel$, which correspond to Bose-Einstein condensates of magnons and are entered via second-order transitions. Based on these numerical results~\cite{Seabra2010}, one may suspect the experimentally observed phases P3 and P7 to be Bose-Einstein condensates.

The additional phases P2, P4, P8, P9, and P10 have no counterparts in the model calculations. Of course, this is not very surprising, because the model used in Refs.~\onlinecite{Seabra2010,Seabra2011a} does not capture the specific aspects of \CsFeBr .
In particular, the $AABB$ stacking of \CsFeBr\ with frustrated antiferromagnetic coupling via $J_c$  may induce additional phases.
%Thus with increasing the longitudinal field \Hvec $\,\parallel\,$\cvec , we can also anticipate the formation of phases in which, {\it e.g.}, not both intra-dimer spins switch simultaneously from up to down, but only one of them.
With increasing longitudinal field \Hvec $\,\parallel\,$\cvec , we can anticipate, e.g., the formation of partial spin-flop phases or incommensurate phases. Indeed, preliminary neutron data indicate that the intermediate zero-field phase P2 is incommensurate and, as shown in Fig.~\ref{MvH_dL_Phadis}, this phase P2 is stabilized for both field directions,  \Hvec $\,\parallel\,$\cvec\ and \Hvec $\,\perp\,$\cvec , but different microscopic spin structures are expected in longitudinal and transverse fields. Thus, further diffraction studies on the field-induced magnetic phases appear very promising to understand this extremely rich phase diagram.

\section{Conclusions}

In summary, we have identified the new material \CsFeBr\ as a frustrated triangular antiferromagnet with surprisingly rich properties. The magnetic ordering occurs with a strong magnetoelastic distortion. In contrast to some Cr-based isostructural materials, the spins of the Fe$_2$Br$_9$ bi-octahedra do not form a dimer singlet ground state, but are in fact ferromagnetically aligned. This agrees with our ab-initio DFT+U calculations, which yield a weak ferromagnetic intradimer coupling $J$ between the Fe spins within the Fe$_2$Br$_9$ bi-octahedra. In contrast, the interdimer coupling $J_p$ within the triangular planes is antiferromagnetic and frustrated. A similar antiferromagnetic exchange is obtained for the interlayer coupling $J_c$ which acts between spins of neighboring $AB$ layers and is frustrated as well. As a consequence, \CsFeBr\ consists of $AABB$-stacked triangular layers with alternating ferromagnetic and antiferromagnetic coupling along \cvec , which adds to the complexity. The magnetic anisotropy $\delta=0.19$ derived from the saturation fields $H_\text{S}^\perp$ and $H_\text{S}^\|$ appears extraordinarily large for spin-5/2 moments of the Fe$^{3+}$ ions with  half-filled $3d$ shells. The strongly different saturation fields are also remarkable because the magnetic susceptibility in the paramagnetic high-temperature phase is essentially isotropic. The origin of the enhanced magnetic anisotropy in the ordered phases is currently unclear. Possibly, it may arise from anisotropic exchange couplings which manifest more strongly in the ordered phases, or the pronounced structural changes upon entering the ordered phase can enhance the single-ion anisotropy. The phase diagram with the magnetic field along the easy axis is very complex and shows a plethora of field-induced phases, which include two phases  with fractional magnetization plateaus, namely $1/2M_\text{S}$ and $1/3M_\text{S}$, and we have indications of at least one incommensurate magnetic phase. Several first-order phase transitions appear with huge hysteresis effects, and sharp lattice deformations occur.  All this makes \CsFeBr\ an extremely interesting material with very rich and unusual properties.

While finalizing this manuscript we became aware of a very recent publication about the closely related material \CsFeCl\ \cite{Ishii2021}. The magnetic phase diagrams derived for this iso-structural material strongly resemble those of Fig.~\ref{MvH_dL_Phadis}, but with reduced $T_{\text{N}}=5.4\,$K, reduced saturation fields $H_\text{S}^\parallel =19.4\,$T, $H_\text{S}^\perp =17.4\,$T and smaller anisotropy $\delta=0.1$. Although the proposed interpretation of Ref.~\onlinecite{Ishii2021} concerning the relative importance of various exchange couplings differs from our conclusions, it is gratifying that the basic experimental features of both materials, \CsFeCl\ and \CsFeBr , are very similar. This confirms that this extremely rich behavior is indeed an intrinsic property of these materials, although the detailed clarification of the magnetic structures of the different field-induced phases requires further studies.

\begin{acknowledgments}
We acknowledge support by the DFG (German Research Foundation) via Project No. 277146847-CRC 1238 (Subprojects A02, B01, and B04),
by the  Bundesministerium  f\"ur  Bildung  und  Forschung,  ProjectNo. 05K19PK1, and by the Ministry of Science and Higher Education of Russia via Project Quantum AAAA-A18-118020190095-4. DFT calculations were performed on the Uran supercomputer at the IMM UB RAS. The neutron data were partly taken on the single crystal diffractometer HEiDi operated jointly by RWTH Aachen University and the J\"{u}lich Centre for Neutron Science (JCNS) within the JARA collaboration. This work was supported by HFML-RU/NWO-I and HLD-HZDR, members of the European Magnetic Field Laboratory (EMFL).
\end{acknowledgments}

\begin{appendix}

\section{Zero-field transitions}
\subsection{Hysteresis and magnetic entropy}
\label{app:HC}

\begin{figure}[b]
	\includegraphics[width=8.5cm]{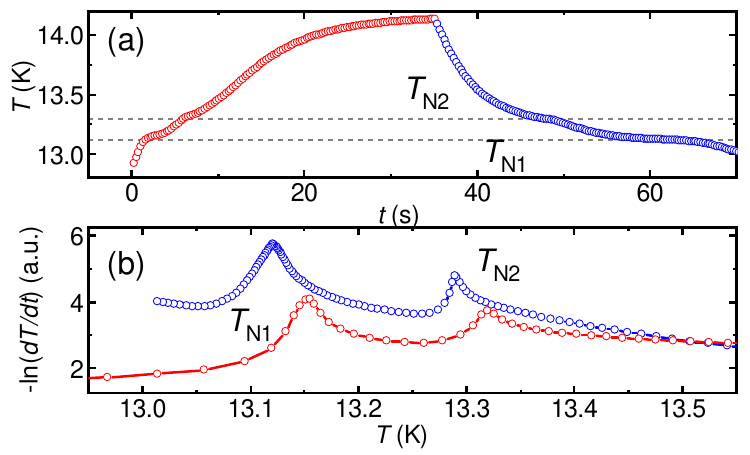}
	\caption{\label{HCpeaks} (a) A long heat pulse in zero field shows two kinks in the time-dependent sample temperature $T(t)$ during  both, the heating run (red) and in the subsequent relaxation curve (blue). The kinks signal $1^{st}$-order transitions at $T_{\text N1}$ and $T_{\text N2}$. (b) A plot of $-\ln\left(\partial T/\partial t \right) \, vs. \, T$ reveals a weak hysteresis of about 30\,mK at each transition.}
\end{figure}

As discussed in Refs.~\onlinecite{LASHLEY2003369,Scheie2018}, the usual relaxation-time method used for specific heat measurements in the PPMS is not well applicable for first-order phase transitions. Thus, long heat pulses over a temperature range of about 1\,K were analyzed. As is shown in Fig.\ref{HCpeaks}(a), the time-dependent evolution of the sample temperature $T(t)$ has pronounced kink in both, the heating run and in the subsequent relaxation curve, which signal two $1^{st}$-order phase transitions with sharp peaks in the heat capacity. The positions of the respective transition temperatures can be obtained from the derivatives $\partial T/\partial t$ as is shown in panel~(b). The transition temperatures $T_{\text N1}$ and $T_{\text N2}$ are separated by 170\,mK and for both transitions, we observe a small hysteresis of 30\,mK between the $T_{\text N}$ values obtained with increasing or decreasing temperature. 

\begin{figure}[t]
	\includegraphics[width=8.5cm]{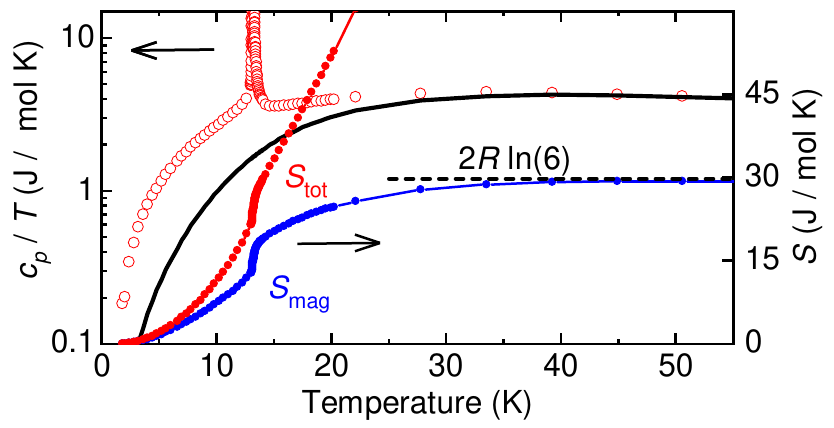}
	\caption{\label{entropy} Heat capacity data ($\circ$; left axis) from Fig.~\ref{chi_cp_dL}(b) in a logarithmic plot of $c_p^{tot}/T \, vs. \, T$  and the corresponding total entropy (red $\bullet$; right axis) obtained by numerical integration. The solid black line is an estimate of the phononic background $c_p^{ph}/T$ that was adjusted such that the magnetic entropy $S_{mag}=\int  (c_p^{tot}-c_p^{ph})/T\, dT$ (blue) approaches $2R \ln(6)=29.8\,$J/molK as expected for \CsFeBr .}
\end{figure}

For the heat capacity data of Fig.~\ref{chi_cp_dL}(b), $c_p(T)$ between 13.2 and 13.4\,K was derived from the relaxation curve of Fig.\ref{HCpeaks} following the procedure described in Ref.~\onlinecite{LASHLEY2003369} and combined with data obtained by the usual relaxation-time method in the remaining temperature ranges. An entropy analysis of these data is presented in Fig.~\ref{entropy}. Temperature intergration of the measured $c_p/T$ data reveals that, despite the rather large $c_p$ anomalies at  $T_{\text N1}$ and $T_{\text N2}$, the combined entropy release at both transitions is about 4\,J/molK. This corresponds to 13\,\% of the full magnetic entropy $2R \ln(2)=29.8\,$J/molK  expected for \CsFeBr\ with two $S=5/2$ moments per formula unit. For conventional magnets, most of the magnetic entropy is expected to change continuously below the ordering temperature, but with decreasing (spin and spatial) dimensionality and/or increasing frustration the continuous magnetic entropy extends towards higher temperature. In order to analyze this, one has to estimate  the phononic background $c_p^{ph}$. Here, we use the solid line in Fig.~\ref{entropy}, which comprises a Debye model with additional Einstein modes. The parameters were adjusted such that (i) $c_p^{ph}(T)$ describes the measured data above about 50\,K and (ii) that the expected $S_{mag}=29.8\,$J/molK is reproduced by $S_{mag}=\int  (c_p^{tot}-c_p^{ph})/T\, dT$. The obtained magnetic entropy release above $T_{\text N1/2}$ appears reasonable in view of the moderate frustration ratio $|\theta|/T_\text{N}\approx4.4$ derived from the Curie-Weiss analysis.

\subsection{Magnetoelastic domains}
\label{app:twins}

Here we consider the occurrence of magnetoelastic domains resulting from the magnetic order of the zero-field phase P1. As shown in Fig.~\ref{magnetic_structure} the in-plane nearest neighbor spins along one of the three originally equivalent hexagonal \avec\ directions are aligned parallel to each other, but alternate along the other two \avec . This results in a 2-fold in-plane symmetry with the orthorhombic $\avec_o$ ($\bvec_o$) axis along (perpendicular to) the line of parallel spins. For a single-domain crystal, different thermal expansion anomalies are naturally expected for $\avec_o$ and $\bvec_o$, if there is a finite magnetoelastic coupling. This can be rationalized by assuming that with the onset of magnetic order the in-plane bond length of the antiparallel spin pairs weakly shrinks while the in-plane bond length of the parallel spin pairs weakly increases.
For a multi-domain crystal, however, the different thermal expansion anomalies of the $\avec_o$ and $\bvec_o$ axes are expected to cancel each other, at least partially. Thus, the rather large and essentially identical $\Delta L_{e_i}/L_{0}$ measured along the different orientations $e_1$ and $e_2$ may appear surprising. However, in the used capacitance dilatometer the crystal is fixed by CuBe springs, which apply a uniaxial pressure parallel to the measured $\Delta L_{e_i}$. This pressure can cause a partial detwinning at the symmetry-breaking phase transition~\cite{Niesen2013, Niesen2014a} and, consequently, the almost identical $\Delta L_{e_i}/L_{0}$ measured along $e_1$ and $e_2$ result from different orientations of twin domains. Note that the dilatometer measurements along $e_1$ and $e_2$ cannot be performed simultaneously.

For a quantitative discussion of the detwinning effects,
% on $\Delta L_i$  measured along $e_1$ and $e_2$
we consider the variations of the magnetoelastic energy $\partial E_\text{mag}/\partial r \propto \delta r$ and of the (nearly) harmonic lattice potential $\partial E_\text{latt}/\partial^2 r \propto \delta r^2$ to lowest order of a generalized lattice distortion $\delta r$. Here, the different powers in $\delta r$ necessarily cause finite lattice distortions in order to minimize the total energy $E_\text{mag}+E_\text{latt}$ and these distortions reflect the magnetic symmetry. Consequently, independent spontaneous strains $\delta a_o$, $\delta b_o$, and $\delta c_o$ along each of the orthorhombic axes are expected for the P1 phase of \CsFeBr . If the variations of $E_\text{mag}$ and $E_\text{latt}$ are restricted to the respective lowest order in $\delta r$, straightforward geometric considerations reveal that $\delta a_o =- \delta b_o$. In this case, the hexagonal-to-orthorhombic transition is area-conserving with respect to the triangular planes, while $\delta c_o$ denotes the bare volume change.
Therefore, no sizeable in-plane length changes should be measured in a fully twinned multi-domain sample, which means that the very large and essentially identical contractions $\Delta L_{e_i}/L_{0}$ measured along the in-plane directions $e_1 \perp e_2$ result from the shape changes of differently oriented twin domains.

\begin{table}[b]
	\caption{\label{tab:table_nuc_refinement} Crystal structure of \CsFeBr \ as refined in space group $P6_3/mmc$ using X-ray data for 150\,K and neutron data
		for 15 and 2.5\,K, respectively. Thermal parameters U$_{ij}$ are given in \AA$^2$
		The structural refinements with the 150\,K X-ray data were performed with the software package \textsc{SHELXL-2018/3} \cite{Sheldrick2015}
		and the neutron data with \textsc{Jana2006} \cite{Petricek2014}. For Br2 also $U_{13}$=2$U_{23}$ can be finite and $U_{23}$ was refined to
		0.0000(2), -0.0002(3) and 0.0011(1)\,\AA$^2$ at 2.5, 15 and 150\,K, respectively. The lattice parameters amount to $a$=7.528(1)\,\AA\ and $c$=18.552(2)\,\AA\ at 150\,K and to $a$=7.491(1)\,\AA\ and $c$=18.477(2)\,\AA\ at 15\,K \cite{Braschoss2019}.}
	\begin{ruledtabular}
		\begin{tabular}{m{0.9cm}ccccc}
			& x  & y  & z &  \\
			& U$_{11}$  & U$_{22}$  & U$_{33}$ & U$_{12}$ \\ \hline
			Cs1&0&0&$1/4$ & {\hfill \hfill $\SI{150}{\kelvin}$} \\
			& 0.0141(4) & =U$_{11}$ & 0.0105(6) &  =0.5U$_{11}$  \\
			Cs2 & $1/3$  & $2/3$ &0.41776(4)  & \\
			& 0.0163(3) &  =U$_{11}$  & 0.0116(4) & =0.5U$_{11}$ \\
			Br1&0.51243(8)  & 2x   &$1/4$ & \\
			& 0.0137(4) & 0.0105(5) & 0.0076(5) & =0.5U$_{22}$ \\
			Br2& 2y & 0.17217(6)  &0.41242(4)  & \\
			& 0.0134(4) & 0.0162(3) & 0.0093(4) & =0.5U$_{11}$  \\
			Fe&$2/3$  & $1/3$ & 0.34667(9)  & \\
			& 0.0140(6) &  =U$_{11}$  & 0.0047(9) & =0.5U$_{11}$ \\
			\multicolumn{6}{m{9cm}}{\centering  R(obs)=4.50,  R(all)=4.74 }\\\hline
			Cs1&0&0&$1/4$ & {\hfill \hfill $\SI{15}{\kelvin}$} \\
			& 0.0030(12) & =U$_{11}$ & 0.009(2) &  =0.5U$_{11}$  \\
			Cs2 & $1/3$  & $2/3$ &0.41779(13)  & \\
			& 0.0060(9) &  =U$_{11}$  & 0.0084(12) & =0.5U$_{11}$ \\
			Br1&0.51173(16)  & 2x   &$1/4$ & \\
			& 0.0047(6) & 0.0063(9) & 0.0079(8) & =0.5U$_{22}$ \\
			Br2& 2y & 0.17137(10)  &0.41275(5)  & \\
			& 0.0046(7) & 0.0056(5) & 0.0091(5) & 0.0023(3) \\
			Fe&$2/3$  & $1/3$ & 0.34658(9)  & \\
			& 0.0038(5) &  =U$_{11}$  & 0.0073(6) & =0.5U$_{11}$ \\
			\multicolumn{6}{m{9cm}}{\centering  R(obs)=3.71, wR(obs)=4.05, R(all)=16.03, wR(all)=4.85}\\\hline
			Cs1&0&0&$1/4$ & {\hfill \hfill $\SI{2.5}{\kelvin}$ }\\
			& 0.0030(12) & =U$_{11}$ & 0.009(2) &  =0.5U$_{11}$  \\
			Cs2 & $1/3$  & $2/3$ &0.41798(8)  & \\
			& 0.0060(9) &  =U$_{11}$  & 0.0084(12) & =0.5U$_{11}$ \\
			Br1&0.51178(11)  & 2x   &$1/4$ & \\
			& 0.0047(6) & 0.0063(9) & 0.0079(8) & =0.5U$_{22}$ \\
			Br2& 2y & 0.17143(7) & 0.41277(4)  & \\
			& 0.0046(7) & 0.0056(5) & 0.0091(5) & 0.0023(3) \\
			Fe&$2/3$  & $1/3$ & 0.34660(6)  & \\
			& 0.0038(5) &  =U$_{11}$  & 0.0073(6) & =0.5U$_{11}$ \\
			\multicolumn{6}{m{9cm}}{\centering  R(obs)=2.44, wR(obs)=2.15, R(all)=4.67, wR(all)=2.44}\\
			
		\end{tabular}
	\end{ruledtabular}
\end{table}

Because $e_1$ and $e_2$ are, respectively, parallel and perpendicular to one of the hexagonal \avec\ axes, these directions correspond to $\avec_o$ and $\bvec_o$ of one domain and are rotated by $\pm 60^\circ$ with respect to the corresponding orthorhombic axes of the other two domains.
If we now consider the case that $\avec_o$ shrinks at the ordering phase transition, the first domain will be favored by uniaxial pressure along $e_1$ and a single-domain state can be reached for large enough pressure. For uniaxial pressure along $e_2$, the other two domains are favored and a single-domain state cannot be reached. When $n_1$ denotes the population of the pressure-induced favored domain, $n_2=1-n_1$ corresponds to the fraction of the other domains, and the total length changes $\Delta L_{e_i}/L_{0}$ measured either along $e_1$ or along $e_2$ are given by
\begin{eqnarray}\label{eq:detwin}
\delta e_1(n_1) &=& \left(\frac{3}{2}n_1-\frac{1}{2}\right)\delta \avec_o \\
\delta e_2(n_1) &=& \left(\frac{3}{2}n_1-\frac{1}{2}\right)\delta \bvec_o=\left(-\frac{3}{2}n_1+\frac{1}{2}\right)\delta \avec_o  \; .
\end{eqnarray}
A fully twinned sample has $n_1=\frac{1}{3}$ resulting in $\delta e_1 = \delta e_2 = 0$. By defining a detwinning ratio $r$ that grows from $0$ for a fully twinned sample to $r=1$ for a single-domain sample, the above equation can be rewritten as
\begin{eqnarray}\label{eq:twinratio1}
\delta e_1(r) &=& \left[\frac{3}{2}\left(\frac{1}{3}+\frac{2}{3}r\right)-\frac{1}{2}\right]\delta \avec_o=r \delta \avec_o\ \\ \label{eq:twinratio2}
\delta e_2(r) &=& \left[\frac{3}{2}\left(\frac{1}{3}-\frac{2}{3}r\right)-\frac{1}{2}\right]\delta \bvec_o=-r \delta \bvec_o %= r \delta \avec_o
\end{eqnarray}
From Eqs.~(\ref{eq:twinratio1}, \ref{eq:twinratio2}), it is clear that, for $\delta \avec_o =-\delta \bvec_o$, the same overall length change can be expected along $e_1$ and $e_2$ if the uniaxial pressure applied either along $e_1$ or $e_2$, respectively, results in the same detwinning ratio $r$. However, for pressure along $e_2$ a maximum detwinning ratio $r=0.5$ can be reached because two domains are equally favorable. As is shown in Fig.\ref{chi_cp_dL}(c), we observed essentially identical contractions along $e_1$ and $e_2$, which restricts the detwinning ratio to $0< r \leq 0.5$ and the intrinsic orthorhombic distortions  correspond to $1/r \cdot \Delta L_{e_i} /L_{0}$. Using the measured $L_{e_i} /L_{0} \simeq 3 \cdot 10^{-4}$  allows us to give  the lower bounds $|\delta \avec_o|,|\delta \bvec_o|  \geq 6 \cdot 10^{-4}$, but we cannot judge which of the orthorhombic axes contracts or expands.

\begin{table}[b]
	\caption{\label{tab:table_irreps} Basis vectors of the irreducible representations obtained for the propagation vector ($\frac{1}{2}$,0,0) in space group
		$P6_3/mmc$ with four magnetic Fe ions at Fe1 ($\frac{2}{3}$,$\frac{1}{3}$,0.35),
		Fe2 ($\frac{1}{3}$,$\frac{2}{3}$,0.85),
		Fe3 ($\frac{1}{3}$,$\frac{2}{3}$,0.65), and
		Fe4 ($\frac{2}{3}$,$\frac{1}{3}$,0.15).     }
	\begin{ruledtabular}
		\begin{tabular}{m{0.9cm}ccccc}
			& Fe1  & Fe2  & Fe3  & Fe4 \\ \hline
			$\Gamma_1$ & $0 \bar{u} 0$ & $0 \bar{u} 0$ & $0 u 0$ & $0 u 0$ \\
			$\Gamma_2$ & $ 2u u v$  & $ 2u u \bar{v}$ & $ 2u u v$ & $ 2u u \bar{v}$ \\
			$\Gamma_3$ & $ 2u u v$  & $ 2u u \bar{v}$ & $ 2\bar{u} \bar{u} \bar{v}$ & $ 2\bar{u} \bar{u}v$ \\
			$\Gamma_4$ & $0 u 0$ & $0 u 0$ & $0 u 0$ & $0 u 0$ \\
			$\Gamma_5$ & $0 u 0$ & $0 \bar{u} 0$ & $0 \bar{u} 0$ & $0 u 0$ \\
			$\Gamma_6$ & $ 2u u v$  & $ 2\bar{u} \bar{u} v$ & $ 2u u v$ & $ 2\bar{u} \bar{u} v$  \\
			$\Gamma_7$ & $ 2u u v$  & $ 2\bar{u} \bar{u} v$ & $ 2\bar{u} \bar{u} \bar{v}$ & $ 2u u \bar{v}$  \\
			$\Gamma_8$ & $0 \bar{u} 0$ & $0 u 0$ & $0 \bar{u} 0$ & $0 u 0$ \\
			
		\end{tabular}
	\end{ruledtabular}
\end{table}

\section{Structural details}
\subsection{Crystal structure}
\label{app:struc}

The crystal structure was analyzed at 150\,K using X-rays~\cite{CSD} and at 15 and 2.5\,K with neutrons.
On the four-circle neutron diffractometer HEiDi nuclear reflections were collected using the wavelengths
1.171\,\AA\ and 0.795\,\AA\ whereas magnetic reflections were only collected with $\lambda$=1.171\,\AA.
Reflections that were corrupted by a varying background or by a too close neighboring Bragg peak were culled manually,
so the following numbers refer to the remaining reflections. For the structural refinement, 1291 reflections were collected
at 15 K (649 of which are unique with respect to space group $P6_3/mmc$ and 282 unique
reflections were observed) and 1109 reflections were collected at 2.5 K (350 of which
are unique and 272 unique reflections were observed).
For the magnetic data collection at 2.5\,K, 336 magnetic reflections were collected
(310 of which belong to different Friedel pairs and 94 of which were observed).

The structural refinements were performed with the software package \textsc{SHELXL-2018/3} \cite{Sheldrick2015} for
the X-ray data taken at 150\,K, and with the \textsc{Jana2006} \cite{Petricek2014} software package for the low-temperature neutron data.
The resulting parameters are given in Table \ref{tab:table_nuc_refinement}. The low-temperature structural parameters are identical within the
error bars for 15 and 2.5\,K but there are slight differences with the results obtained at 150\,K.
The largest deviations can be found in the x and y coordinates of the Br atoms.

\begin{figure}[t]
	\includegraphics[width=8.cm]{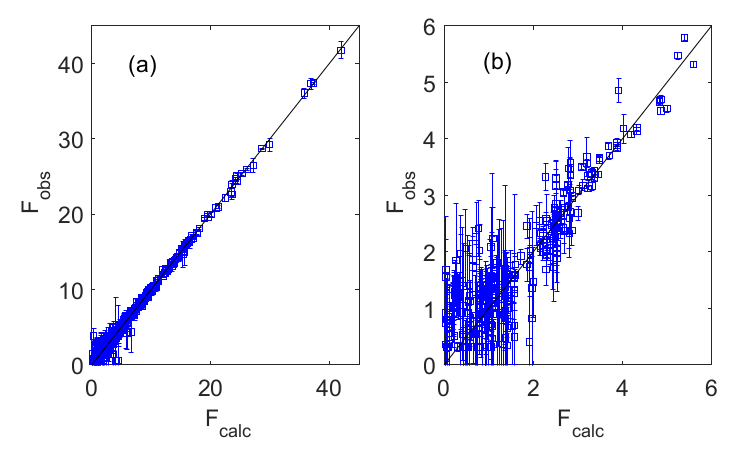}
	\caption{(a, b) The observed structure factors are plotted against the calculated ones for the refinements of the nuclear and magnetic structure with
		the nuclear and magnetic Bragg reflection intensities, respectively (T=2.5\,K). The Fullprof program package \cite{Rodriguez-Carvajal1993} was used in these refinements.}
	\label{fobs-fcalc}
\end{figure}

\subsection{Magnetic structure}
\label{app:mag}

The symmetry analysis of the zero-field magnetic structure was performed with the Fullprof program package \cite{Rodriguez-Carvajal1993} and is presented in Table~\ref{tab:table_irreps}. Fig. \ref{fobs-fcalc} presents the comparison of observed and calculated structure factors for the nuclear and magnetic reflections in panel (a) and (b), respectively.

\begin{figure}[b]
	\includegraphics[width=6.5cm]{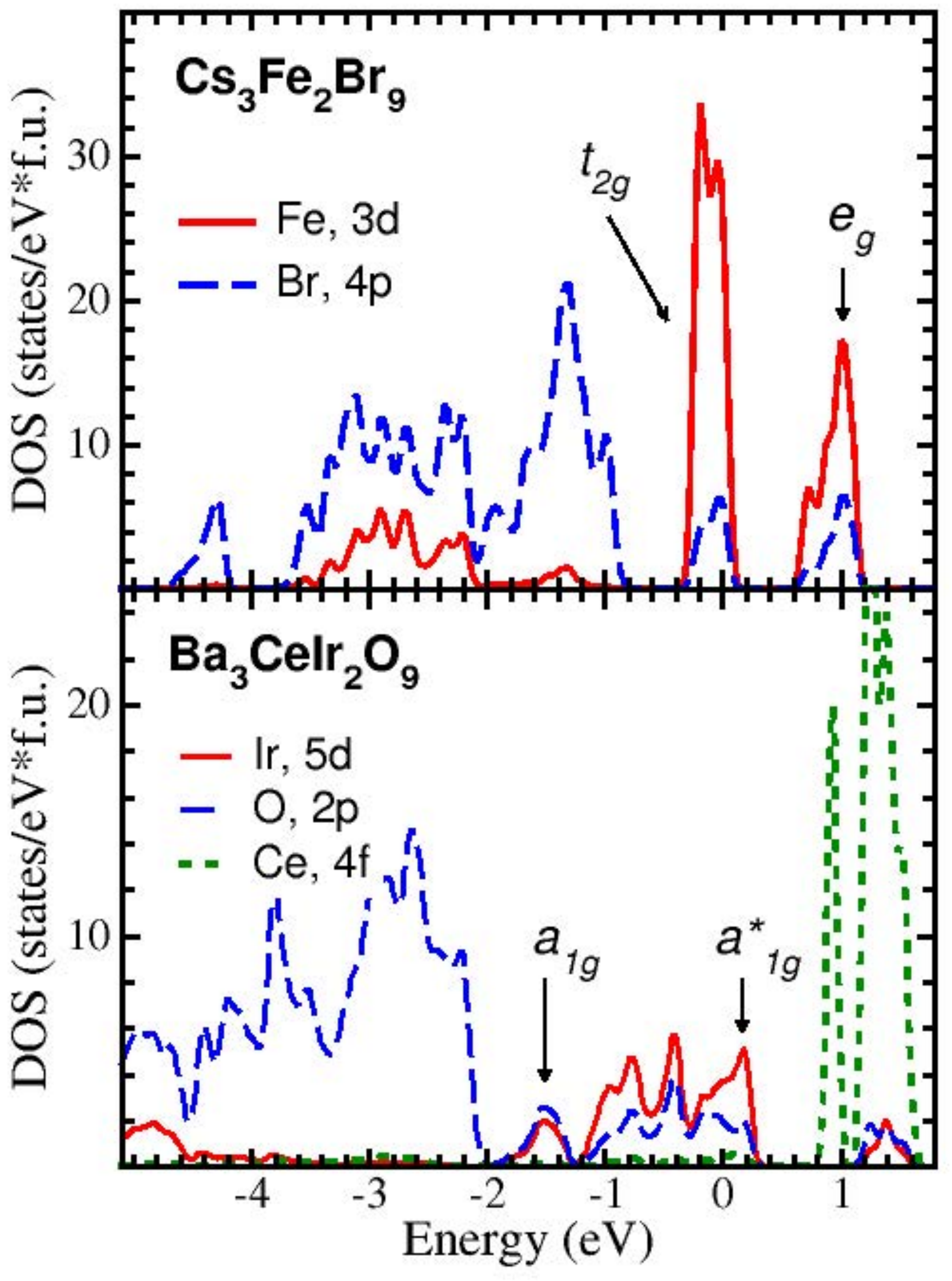}
	\caption{\label{DOS} Partial density of states (DOS) of Cs$_3$Fe$_2$Br$_9$ in comparison to Ba$_3$CeIr$_2$O$_9$ (also having structural dimers with face-sharing bi-octahedra) as obtained in non-magnetic GGA calculations. Molecular Ir-Ir orbitals are seen in the lower panel ($a_{1g}$ marks bonding and $a_{1g}^*$ antibonding
    states of the corresponding symmetry) in contrast to the ordinary atomic $t_{2g}$  and $e_g$ orbitals in case of  Cs$_3$Fe$_2$Br$_9$. Fermi energy is zero.}
\end{figure}

\section{Details of DFT+U calculations}
\label{app:DFT}

All calculations were performed using the full-potential linearized augmented plane-wave method as realized in the wien2k code~\cite{w2k}. We used the Perdew-Burke-Ernzerhof version of exchange-correlation functional~\cite{PBE}. The supercell used for the total energy calculations consisted of four formula units. The irreducible part of the Brillouin zone was sampled by a 9$\times$5$\times$9 $k$-point grid. Strong Coulomb correlations were taken into account using the DFT+U method~\cite{Licht}, and Hubbard on-site electron repulsion and Hund's intra-atomic exchange parameters were chosen to be $U=5.8$ and $6.5\,$eV and $J_H=0.95\,$eV as for other similar materials~\cite{FeO2,FeS}.

The upper panel of Fig.~\ref{DOS} illustrates results of non-magnetic GGA calculations of \CsFeBr . One can see that the electronic structure in this case is very different from what we have in another dimer-material Ba$_3$CeIr$_2$O$_9$ with a very similar crystal structure~\cite{Revelli2019},  whose density of states are presented in the lower part of Fig.~\ref{DOS}. In Ba$_3$CeIr$_2$O$_9$ one clearly sees the formation of the bonding ($a_{1g}$) and antibonding ($a_{1g}^*$) bands. There is nothing like this in \CsFeBr , where one may distinguish the atomic $t_{2g}$ and $e_g$ states only. Thus, in spite of naive expectations \CsFeBr\ should not be considered as a material with molecular orbitals formed by the Fe-$3d$ states and a there is no large exchange coupling between sites as a result. The reason for this is a large ionic radius of Br$^{1-}$ (1.96~\AA) compared to O$^{2-}$ (1.4~\AA)~\cite{Shannon:a12967}, which results in a large a interatomic distance of $\simeq 3.6\,$\AA\ between the Fe$^{3+}$ ions.

\end{appendix}

%\bibliography{Cs3Fe2Br9_2021,Cs3Fe2Br9_2021_add}
%\end{document}

%apsrev4-2.bst 2019-01-14 (MD) hand-edited version of apsrev4-1.bst
%Control: key (0)
%Control: author (8) initials jnrlst
%Control: editor formatted (1) identically to author
%Control: production of article title (0) allowed
%Control: page (0) single
%Control: year (1) truncated
%Control: production of eprint (0) enabled
%

\end{document}